\documentstyle[preprint,prb,aps]{revtex}

\begin{document}
\draft
\preprint{}
\title{Random Exchange Disorder \\
in the Spin-1/2 XXZ Chain
}
\author{Stephan Haas, Jose Riera\cite{byline}, and Elbio Dagotto}
\address{
Department of Physics and Supercomputer Computations Research
Institute,\\ Florida State University,
Tallahassee, Florida 32306}
\date{\today}
\maketitle
\begin{abstract}
The one-dimensional XXZ model is studied in the presence of disorder
in the Heisenberg Exchange Integral. Recent predictions obtained from
renormalization group calculations are investigated numerically using a
Lanczos algorithm on chains of up to 18 sites. It is found that
in the presence of strong X-Y-symmetric random exchange couplings, a
``random singlet'' phase with quasi-long-range order in the
spin-spin correlations persists.
As the planar anisotropy is varied, the full zero-temperature phase
diagram is obtained and compared with predictions of Doty and Fisher
[Phys. Rev. B {\bf 45 }, 2167 (1992)].
\end{abstract}

\pacs{PACS Indices: 64.10.Pf, 05.30.Jp, 05.70.Jk}

\narrowtext
The study of quantum models in the presence of disorder is an emerging
field onto which much attention has been focussed lately. Since all
experimentally accessible systems \cite{wu} are to some
extent affected by randomness in the form of impurities, fields, or
couplings, a thorough understanding of disorder effects can help in
comparing experimental observations and theoretical predictions.
In particular, weakly disordered, low-dimensional quantum spin systems are of
interest, since the interplay between randomness and
strong quantum fluctuations can be
observed.\cite{oldwork}
At T=0, phase transitions in quantum spin models are driven by zero-point
fluctuations, as opposed to thermal excitations in their classical
counterparts.
However, when a random potential is introduced, phase transitions can be also
driven by random fluctuations.
This mechanism is particularly interesting in the case of marginally ordered
systems, where the long-range N\'{e}el order in the 2D isotropic Heisenberg
model has been found to be
unstable towards thermal fluctuations and random fields, but not towards
randomness in the exchange couplings.\cite{murthy}

The anisotropic spin-1/2 Antiferromagnetic Heisenberg Chain is a generic
model of strongly correlated electrons. It is described by the
Hamiltonian,
\begin{equation}
H_0 = J \sum_i (\lambda S^z_i S^z_{i+1} + S^x_i S^x_{i+1} + S^y_i S^y_{i+1}),
\end{equation}
where the notation is standard. Due to the low
dimensionality, quantum fluctuations destroy long-range order in the region
$-1 < \lambda \leq 1$, and the spin-spin correlations decay
spatially with a power-law.
Beyond the Heisenberg point (i.e. $\lambda > 1$), a gap
opens in the excitation spectrum and the system develops long-range N\'{e}el
order with exponentially decaying correlation
functions,
while for $\lambda \leq -1$ there is a ferromagnetic region with Ising-type
long-range order.

Let us now introduce disorder in the form of X-Y symmetric random exchange
couplings, i.e. such that the planar symmetry of $H_0$ is not broken by the
random
potential,
\begin{equation}
H_{random} = \sum_i \delta_i (S^x_i S^x_{i+1} + S^y_i S^y_{i+1}).
\end{equation}
The random couplings $\delta_i$ are drawn from a uniform distribution
$P(\delta_i) = \theta (\delta_i - \delta J_{xy})\theta (\delta J_{xy} -
\delta_i)$,
where $\langle \delta_i \rangle = 0$ and
$\langle (\delta_i)^2 \rangle = 2( \delta J_{xy} )^3/3$.
The cut-off parameter
$\delta J_{xy}$ serves as a measure for the strength of the random potential.
The physical properties induced by this distribution are believed to be
universal. However, in order to
test this idea, we also studied random exchange couplings drawn
from a Gaussian distribution,
$P(\delta_i) = \frac{1}{\sqrt{2 \pi \sigma_{xy}}} \exp{(- \delta_i^2 / 2
\sigma_{xy})}$.
Here, $\sigma_{xy}$ serves as a measure of the random strength.

The properties of XXZ chains in the presence of various random potentials have
recently been studied by C. A. Doty and D. S. Fisher using renormalization
group
techniques.\cite{doty} It was found that, while random transverse
fields destroy the
(quasi)-long-range spin order, a power-law decay of the spin correlations
may persist in the presence
of random exchange couplings as long as the random Hamiltonian does not break
the planar symmetry of $H_0$.
In particular, it was predicted that a quasi-long-range-ordered phase
extends from the X-Y regime ($-1 < \lambda \leq 1$), when $H_{random}$ is
switched on.

In our study of the above system, we numerically diagonalized chains of up
to 18 sites with periodic boundary conditions using a Lanczos algorithm.
The observables were obtained from a quenched average, i.e. the ground state
$|\phi_0 (j) \rangle$ of a chain was obtained for a given set  of
random couplings $j = \{ \delta_i \}$, and
then the expectation value of some particular operators $\hat{O}$
were studied.
This procedure was repeated for m $\simeq 500$ different sets of random
couplings, and
finally the algebraic average over all m random samples was taken.
The quenched average of an operator $\hat{O}$ is thus defined by
\begin{equation}
\langle \langle \hat{O} \rangle \rangle = \frac{1}{m} \sum_{j=1}^m \langle
\phi_0 (j)
|\hat{O} |\phi_0 (j) \rangle.
\end{equation}

First, we would like to address the question of whether quasi-long-range order
persists in the region $-1 < \lambda \leq 1$ when the disorder potential
$H_{random}$ is switched on. The relevant observable is the normalized
real-space spin-spin correlation function
\begin{equation}
\omega^z (l) = \frac{3}{N} \sum_{i=1}^N \frac{\langle \langle S^z_i
S^z_{i+l} \rangle \rangle}
{S(S+1)},
\end{equation}
where N denotes the number of sites, and $S=1/2$ in our study.

In Fig.1, the spin-spin correlations $\omega^z (N/2)$ at the maximum separation
($l=N/2$) are plotted as a function of the lattice
size N at planar anisotropy $\lambda = 0.5$ for a couple of random strengths
$\delta J_{xy}$. If the correlations decay with a power-law $|\omega^z(l) |
\propto l^{- \eta_z}$, we expect a straight line with negative slope $\eta_z$
in a double-logarithmic plot. It is found that for all random
strengths, $\delta J_{xy}$, a power-law decay (solid line) fits the numerical
data much better than an exponential decay (dashed line),
e.g. the $\chi^2$-value obtained from least-square fits is typically
two orders of magnitude larger when an exponential decay $|\omega^z(l) |
\propto \exp{(- \xi l)}$ is assumed. We observed a similar power-law
behavior in a large region of parameter space.

Why does the random potential not destroy
quasi-long-range order in this region?
According to Doty and Fisher the ``random singlet" phase which extends from
the X-Y phase of the pure system ($H_0$) can be pictured
in terms of randomly distributed tightly coupled singlet pairs of
spins.\cite{doty} Those spins which are not bound in
a singlet pair interact via virtual excitations. It turns out that these
``almost-free" spins are anomalously strongly correlated. The probability
that ``almost-free" spins separated by a distance $R$ interact strongly is
proportional to $1/R^2$.
This gives rise to the observed power-law behavior in the spin-spin
correlations.
The decay exponent is found to be $\eta_z =2$.
Note that also, in the exactly solvable X-Y limit ($\lambda = 0$) the system
maps into a tight-binding model of free fermions with random nearest-neighbor
hopping. In this limit the decay exponent is
given by $\eta_z =2$ if a single characteristic localization length is
assumed for the properties of the low-energy wave functions.\cite{doty}

In Fig.2, we show $\eta_z$ obtained in our numerical analysis,
as a function of the disorder parameter
$\delta J_{xy}$ for various anisotropies $\lambda$. The exponent has been
extracted using chains of size N=6, 10, 14 and 18.
\cite{scalinglaw}
The inset of Fig.2(a)
shows the decay exponent $\eta_z$ for the pure system $H_0$ as it has also been
obtained in Ref. 6. The exact diagonalization results are in excellent
agreement with
predictions from conformal invariance,\cite{alcaraz} and in particular the
Heisenberg limit ($|\omega^z(l) | \propto l^{-1}$) and the X-Y limit
($|\omega^z(l) |\propto l^{-2}$) are nicely recovered.
For negative anisotropies,
($-1 < \lambda \leq 0$) conformal invariance predicts a constant exponent
$\eta_z =2$, which is also in reasonable agreement with our data,
showing that our techniques can reproduce known results very accurately.

On our finite chains and
as we depart from the $\delta J_{xy} =0$ limit, three regions can be
identified:

(1) In the regime of small randomness ($\delta J_{xy} < J$) the exponent
$\eta_z $ increases slightly as a function of the disorder parameter
$\delta J_{xy} $, which is a sign of reduced order.

(2) Around  $\delta J_{xy} = J$, there is an area of high competition
between the quantum fluctuations of the original Hamiltonian
($J \sum_i (S^x_i S^x_{i+1} + S^y_i S^y_{i+1}$) and $H_{random}$.
Locally the random terms can compensate the zero-point fluctuations leading
to an antiferromagnetic
Ising-like behavior in the correlation functions. As a result, the
decay exponent $\eta_z$ has a dent with onset at around $\delta J_{xy} = J$,
indicating a crossover into a more ordered Ising-like regime, where
correlations decay more slowly than for the uniform system.

(3) For large disorder, ($\delta J_{xy} >> J$) $H_{random}$ is the dominant
term. The dependence of the decay exponent on the planar anisotropy in
$H_0$ becomes negligible, and it approaches $\eta_z =2$ for all values
of $\lambda$, as it has been
predicted by renormalization group arguments.\cite{doty}

In the vicinity of $\lambda = -0.75$ the exponent $\eta_z$ behaves
anomalously for small disorder. The observed decay in
$\eta_z$ for $\delta J_{xy}$ between $J$ and $2J$ is due to ferromagnetic
behavior in the real space spin-spin correlations.
This anomaly is observed specially
for anisotropies
$-1 < \lambda \leq -0.5$.
The dent of $\eta_z$ around $\delta J_{xy} = J$ can be understood as
a crossover into a phase of higher order. In particular, for
$\lambda = -0.75$ we observed a transition into a partially polarized phase
indicated by the change of sign in the energy difference
$\delta E = E(S^z_{tot} = 0) - E(S^z_{tot} = 1)$, where $E(S^z_{tot} = n
)$
is the quenched ground state energy in the subspace with $S^z_{tot} = n$.
The inset of Fig.2(b) shows $\delta E$ as a function of the disorder
parameter $\delta J_{xy}$ at anisotropy $\lambda = -0.75$ for a 14-site
chain. It can be nicely seen that the transition into the partially
polarized phase ($0.55J \leq \delta J_{xy} \leq 3.05J$) corresponds
to the dent in $\eta_z$ in the same regime of disorder.

In Fig.3(a), the dependence of the energy on the disorder parameters
$\delta J_{xy}$ and $\sigma_{xy}$ at various anisotropies is shown for a
14-site chain. As the random potential becomes dominant, the system is
allowed to relax into a ground state of higher entropy. The ground state
energy drops proportionally to $\delta J_{xy}$ ($\sigma_{xy}$) in this
region. In Fig.3(b), we show how the static structure factor
($S^{zz}(k) = \sum_j \exp (-i k j) \langle 0
| S^{z}_j S^{z}_{j+1} | 0 \rangle$) behaves as a function of the disorder
parameters at antiferromagnetic momentum transfer $k = \pi$ for the 14-site
chain. In analogy to Fig.2, three regions can be identified. At low disorder
the structure factor remains approximately unchanged. In the region of
competition,
N\'{e}el order is favored for positive anisotropies ($0 \leq \lambda \leq 1$),
resulting in an increase of
the antiferromagnetic structure factor especially in the vicinity of
$\lambda \sim 1$. For negative anisotropies ($-1 < \lambda \leq 0$),
the ditch in $S^{zz}(\pi)$ indicates a crossover into a ferromagnetically
polarized region. For large disorder, $S^{zz}(\pi)$ becomes independent of
$\lambda$, and approaches the X-Y limit for all anisotropies.

The boundary between the long-range-ordered regime and the ``random singlet"
phase
is obtained from the correlations $\omega^z(N/2)$.
In the ``random singlet" phase, the spin-spin correlations at distance N vanish
in the bulk limit as ${\bf N \rightarrow \infty}$. However, as the anisotropy
is tuned
across the critical value $\lambda_c$, $\omega^z(N/2)$ becomes finite,
approaching $| \omega^z(N/2) | =1 $ in the extreme Ising limit ($\lambda =
\infty$).
At zero disorder the Heisenberg point $\lambda_c = 1$ is nicely recovered as
the critical point (Fig.3(c)).
In Fig.3(d), we see that the transition point between these two phases
is $reduced$ to about $\lambda_c = 0.75$ at $\delta J_{xy} = J$.
\cite{intermediate}
As a result of the strong competition effects in the
region $\delta J_{xy} \simeq J$, the antiferromagnetic phase bends into the
random singlet regime in a $``reentrant"$
transition, indicating a stronger antiferromagnetic order in this region.
The whole boundary between ``random singlet" and N\'{e}el phase is plotted in
the phase diagram given in Fig.4.

Both the ``random singlet" and the N\'{e}el phase lie in the $S^z_{total} =0$
subspace. On the other hand, as
the ferromagnetic limit is approached, there is a transition into
a partially polarized phase, i.e. the ground state no longer has
$S^z_{total} =0$. This phase boundary, as well as the transition
from the partially into
the fully polarized regime, is extracted from comparing the lowest energies
of the various $S^z_{total}$ subspaces (averaged over the ensemble of
random couplings).
In the region of competition
between quantum fluctuations and the disorder term, the partially polarized
phase bends into the ``random singlet" regime, in analogy to the effect at the
phase boundary between the ``random singlet" and the N\'{e}el phase, as
shown
in Fig.4.

For low disorder, our results agree qualitatively with those of Doty and
Fisher.\cite{doty} However, their study predicts an X-Y-like ``mole hill"
phase in the region $-1 < \lambda \leq -0.5$, and for small disorder.
Numerically,
it is hard to distinguish this ``mole-hill" from the ``random singlet" regime,
because both phases show power-law behavior in the correlation functions.
However, from our exact diagonalization data we have observed a region
(denoted with a question mark in Fig.4) which has power-law decay, and is a
member of the $S^z_{total} =0$ subspace, but does not have any remnant
antiferromagnetic correlations, as has been discussed above
in the inset of Fig.2(b).
We are currently investigating, whether
this regime can be identified with the ``mole hill" predicted by Doty
and Fisher.

In summary, we have presented the first numerical study of
the spin-1/2 XXZ chain in the presence
of a random exchange potential ($H_{random}$).
In contrast to a random field,\cite{zimanyi}
quasi-long-range order of the zero-disorder X-Y regime $-1 < \lambda \leq 1$
is not destroyed by an X-Y symmetric random exchange. Also, Ising-type
long-range order persists in the presence of small random exchange couplings.
The power-law behavior in the ``random singlet" phase may be due to virtual
interactions of ``almost-free" spins which are not bound in randomly
strong singlet pairs. A complete phase diagram is provided. In addition,
we have found an interesting reentrant transition of the ordered phases
(in both the ferromagnetic and
antiferromagnetic Heisenberg limits) when exchange disorder is included. Such a
novel
type of behavior (order induced by random couplings) deserves additional study.

We thank A. Moreo, F. Alcaraz, T. Barnes, G. Zimanyi,
and K. Runge for useful discussions.
J.R. has been supported in part by the U. S. Department
of Energy (DOE) Office of Scientific Computing under the
High Performance Computing and Communications Program (HPCC),
and in part by DOE under contract No. DE-AC05-84OR21400 managed by
Martin Marietta Energy Systems, Inc., and under contract No.
DE-FG05-87ER40376 with Vanderbilt University.
The work of E. D. is partially supported by the ONR grant
N00014-93-1-0495, and by the donors of The Petroleum Research Fund.
The support of the Supercomputer Computations Research Institute
(SCRI) is acknowledged. The computer calculations were carried out
on the CRAY-YMP at Florida State University.

         \begin{figure}
\caption{: Double-logarithmic plot of real-space spin-spin correlations
$| \omega^z (l) |$ at
maximum separation ($l=N/2$)
as a function of lattice size. The squares represent
data obtained from exact diagonalizations, the solid lines are fits to
power-law decay $|\omega^z(l) |= A l^{-\eta_z}$ and the dashed lines are fits
to an exponential decay $|\omega^z(l) |= A \exp{(- \xi l)}$. The size of the
squares is comparable to the magnitude of the corresponding error bars.}
         \end{figure}

          \begin{figure}
\caption{: (a) Exponents of the power-law decay
$|\omega^z(l) |= A l^{-\eta_z}$
as a function of the disorder parameter $\delta J_{xy}$ for various
positive planar anisotropies.
The inset shows $\eta_z$ as a function of anisotropy
in the limit of no disorder.
(b) Same as (a) but for negative anisotropies.
The inset shows the energy difference $\delta E = E(S^z_{tot} = 0) -
E(S^z_{tot} = 1)$ as a function of $\delta J_{xy}$ for the 14-site chain.
The change in the sign of $\delta E$ indicates
the presence of a partially polarized phase for
$0.55J \leq \delta J_{xy} \leq 3.05J$ at anisotropy $\lambda = -0.75$. }
            \end{figure}

\begin{figure}
\caption{: (a) Ground state energy of the 14-site spin-1/2
XXZ chain as a function of the disorder parameter
$\delta J_{xy}$ at various planar anisotropies. The random exchange couplings
are drawn from a uniform distribution with cut-off $\delta J_{xy}$.
The inset shows the same except
when the random exchange couplings are obtained from a Gaussian distribution
of width $\sigma_{xy}$.
(b) Antiferromagnetic structure factor vs. disorder parameter $\delta J_{xy}$
for the 14-site spin-1/2 XXZ chain.
The inset shows the same but
when the random exchange couplings are obtained from a Gaussian distribution
of width $\sigma_{xy}$.
(c) Real space correlation functions at the maximum distance for an $N$
site chain
as a function of anisotropy at zero
disorder. The bulk limit $N= \infty$ is extracted from a finite size study.
(d) Same as (c) at disorder $\delta J_{xy} = J$.}
\end{figure}

\begin{figure}
\caption{: Phase diagram of the spin-1/2 XXZ chain in the presence of
a random exchange potential. The question mark denotes the ``mole hill"
phase discussed in the text.}
\end{figure}


\begin{references}
\bibitem{scalinglaw} The exponent $\eta_z$ was extracted from finite size
data for the correlation function at maximum separation $\omega^z (N/2)$
using $\eta_z (N/2) \sim
\frac{\ln (A/\omega^z (N/2))}{\ln (N/2)}$ at large $N$.
\bibitem{intermediate} Since the $\omega^z (N/2)$ were measured only at
certain anisotropies ($\lambda = 0.0,0.5,1.0, ...$), the critical
anisotropies $\lambda_c$ were roughly obtained from a linear extrapolation
using $\omega^z (\infty )$ for $\lambda$ values closely above the point
where the spin gap opens.
\end{references}
\end{document}